\documentclass[11pt,twoside]{article}

\usepackage{asp2014}

\aspSuppressVolSlug
\resetcounters

\begin{document}

\title{HSTCosmicrays: A Python Package for Analyzing Cosmic Rays in HST Calibration Data}

\author{N. D. Miles,$^1$ S. Deustua,$^1$ and G. Tancredi$^2$}
\affil{$^1$Space Telescope Science Institute, Baltimore, Maryland, USA; \email{nmiles@stsci.edu}, \email{deustua@stsci.edu}}
\affil{$^2$Universidad de la Rep\'{u}blica de Uruguay, Montevideo, Uruguay; \email{gonzalo@fisica.edu.uy}}

\paperauthor{Nathan Miles}{nmiles@stsci.edu}{}{Space Telescope Science Institute}{}{Baltimore}{Maryland}{21218}{USA}
\paperauthor{Susana Deusta}{deustua@stsci.edu}{}{Space Telescope Science Institute}{}{Baltimore}{Maryland}{21218}{USA}
\paperauthor{Gonzalo Tancredi}{gonzalo@fisica.edu.uy}{}{Universidad de la Rep\'{u}blica de Uruguay}{}{Montevideo}{}{}{Uruguay}



\begin{abstract}
\texttt{HSTCosmicrays} is a python-based pipeline designed to find and characterize cosmic rays found in dark frames (exposures taken with the shutter closed).  Dark exposures are obtained routinely by all the Hubble Space Telescope (HST) instruments for calibration. The main processing pipeline runs locally or in the cloud on AWS . To date, we have characterized more than 1.2 billion cosmic rays in $\sim$76,000 dark frames obtained with CCDs from the four active instruments ACS/HRC, ACS/WFC, STIS, WFC3/UVIS, and the legacy instrument WFPC2. 

\end{abstract}




\section{Introduction}
The Hubble Space Telescope (HST) has been operating outside the terrestrial atmosphere in low earth orbit (LEO) since 1990. At these altitudes, every image obtained by a solid state detector (e.g. CCD) is contaminated by high energy particles (hereafter referred to as cosmic rays). These cosmic rays (electrons, protons and atomic nuclei) originate from within our galaxy in supernovae explosions and from within our solar system in coronal mass ejections and solar wind. The flux observed by HST encodes information about their interaction with the Earth's magnetic field in LEO. By studying these cosmic ray events, as well as the rates at which they occur, we can develop more robust cosmic ray rejection routines and examine the interaction of the Earth's magnetic field with the interplanetary field over time.
 
We have written the python package, \texttt{HSTCosmicrays}, to identify and analyze cosmic rays in dark frames (images taken with the shutter closed) taken as part of routine calibration programs for instruments onboard HST. In Section \ref{s:pipeline}, we provide a high level overview of the package. In Section \ref{s:identification}, we detail the processes by which we identify and label cosmic rays. In Section \ref{s:analysis}, we detail the data extracted by the pipeline. Finally, in Section \ref{s:results}, we detail a key result obtained using data extracted by the pipeline.

\section{The Pipeline}\label{s:pipeline}
 The pipeline is broken into 5 distinct steps that are applied to consecutive one month intervals and all the darks they contain; download, process, analyze, save, and clean up. It is cloud enabled and can be run on AWS utilizing the HST Public Dataset. For each month, we use \texttt{astroquery} \citep{astroquery} to programmatically query and download the bias-corrected dark frames and their engineering and telemetry files from the Mikulski Archive for Space Telescopes (MAST). After downloading, the images are processed and then analyzed to extract statistics about each cosmic ray identified in the image. Once the analysis has completed, the results are written to file and an email is sent to a user defined email address with summary statistics of all the cosmic rays found in each image. Finally, all downloaded images and any temporary files created along the way are deleted to prepare for the next one month chunk of darks.  

\section{Cosmic Ray Identification}\label{s:identification}

For the active instruments (ACS, STIS, WFC3), we leverage each instrument's cosmic ray rejection routine to process each month of images. These routines implement a noise-based rejection model that looks for statistically significant outliers in CR-SPLIT observations taken with HST, i.e. observations made at the same exact pointing in sequence. Each pixel identified by the algorithm is marked with a special bit flag, 8192, in the data quality (\texttt{DQ}) array of the multi-extension FITS file.  A bitwise-and comparison is performed to generate a binary image where pixels affected by cosmic rays have a value of 8192 and everything else 0.

For the legacy instruments (WFPC2), a different method is used to avoid the IRAF dependency of the calibration software. First, we perform three iterations of sigma clipping to compute an accurate estimation of the image statistics (mean, median, and standard deviation). Next, using the sigma-clipped mean and standard deviation we create a binary image by identifying all pixels that are more than three standard deviations above the mean. 

\begin{figure}[ht] 
   \centering
   \includegraphics[scale=0.8]{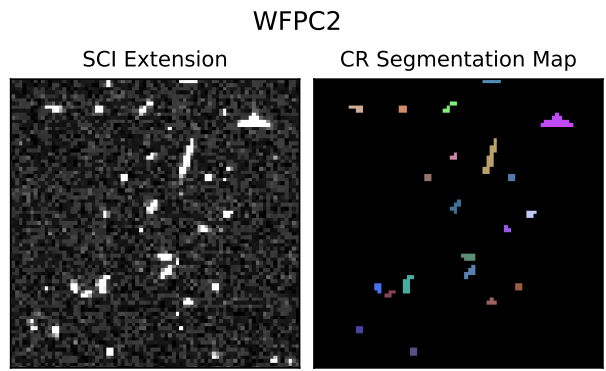} 
   \caption{An example of the final label generated using a random dark frame taken with WFPC2.}
   \label{fig:cr_label}
\end{figure}

In each case, after the binary image has been created we perform a connected-component labeling analysis using the 8-connectivity matrix. The labeling analysis identifies groups of pixels affected by the same cosmic ray. To minimize contamination from hot pixels, we reject any objects identified that affect 1 or 2 pixels. The end result is a segmentation map of the uniquely identified cosmic rays (Figure \ref{fig:cr_label}).

\section{Cosmic Ray Analysis}\label{s:analysis}
\begin{figure}[ht]
   \centering
   \includegraphics[scale=0.6]{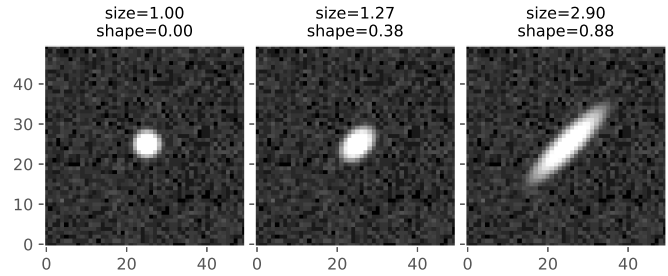} 
   \vspace{-0.5cm}
   \caption{Simulated sources with their corresponding size and shape measurements denoted at the top of each subplot.}
   \label{fig:example_morph}
\end{figure}

Once the cosmic ray label has been generated, we apply the label to the \texttt{SCI} extension to derive quantities of interest describing each identified cosmic ray. First, we compute the total number of pixels affected (i.e. the size in attached pixels) and record the coordinates of each pixel affected. Next, we use the label to compute the moments of the energy distribution,
\begin{enumerate}
\item $I_0 = \sum_{i} p_i$
\item $I_x = \frac{1}{I_0} \sum_{i}p_i * x_i $
\item $I_y = \frac{1}{I_0} \sum_{i} p_i * y_i $
\item $I_{xx} = \frac{1}{I_0} \sum_{i}p_i(x_i - I_x)^2$
\item $I_{yy} = \frac{1}{I_0} \sum_{i}p_i(y_i - I_y)^2$
\item $I_{xy} = \frac{1}{I_0} \sum_{i}p_i(x_i - I_x)*(y_i - I_y)$
\end{enumerate}
where $p_i$ is the pixel value of the $i^{th}$ pixel in the cosmic ray label and $x_i$, $y_i$ are the x and y coordinates of the $i^{th}$ pixel, respectively. The first parameter is the total energy deposited in the detector by the cosmic ray. The second and third parameters combine to give the centroid as $(I_x, I_y)$. The fourth and fifth parameters are used to compute a measure of width or "size" as,
\begin{equation}
\sqrt{\frac{I_{xx} + I_{yy}}{2}}.
\end{equation}
The fourth, fifth, and sixth parameters are used to asses the shape or symmetry of the energy distribution,
\begin{equation}
\sqrt{\frac{(I_{xx} - I_{yy})^2 + 4I^2_{xy}}{(I_{xx} + I_{yy})^2}}.
\end{equation}
In Figure \ref{fig:example_morph}, we show simulated sources over a poisson background to visually demonstrate the size and shape parameters. For each source, we use a 2D Gaussian distribution, 
\begin{equation}
	p(x, y) = A*e^{-\left({\frac{(x - x_0)^2}{\sqrt{2\sigma_x}} + \frac{(y - y_0)^2}{\sqrt{2\sigma_y}}}\right)},
\end{equation}
with $(x_0, y_0)=(25, 25)$,  $A=10^4$, and an added rotation of $\pi/4$. The source on the left is the ideal, perfectly symmetric 2D Gaussian with  $\sigma_x=\sigma_y=1$. The middle source has been skewed such that $\sigma_y=1.5*\sigma_x$. The source on the right has been skewed such that $\sigma_y=4*\sigma_x$.

For each image, we compute the cosmic ray rate, defined as the total number of cosmic rays divided by the total integration time (exposure time plus half of the readout time) and the geometric size of the detector. Next, for each cosmic ray identified, we record the coordinates of each pixel affected during the observation. Finally, in addition to the morphological parameters, we store the following metadata for each image: altitude, latitude, longitude, observation date, observation start time, observation end time, the total integration time, and the WCS information.

\vspace{-0.75cm}
\section{Results}\label{s:results}
\begin{figure}[ht] 
   \centering
   \includegraphics[scale=0.65]{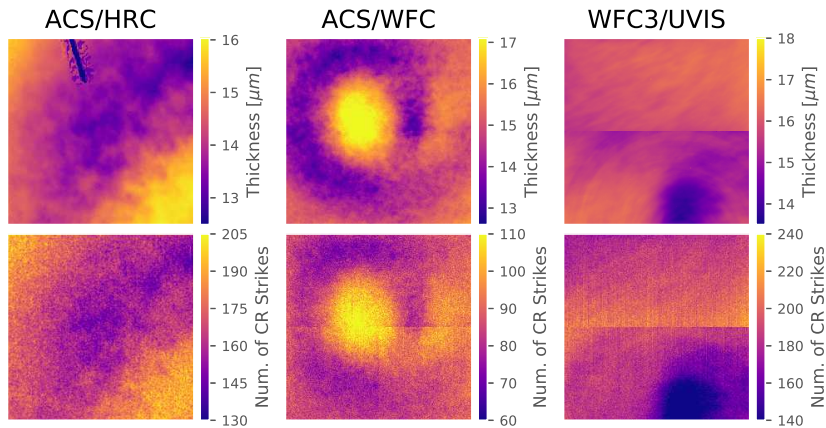} 
   \vspace{-0.45cm}
   \caption{A comparison of detector thicknesses and cosmic ray incidence maps. As expected, thicker regions of the detectors are impacted more often. The ACS and WFC3 thickness maps were provided by J.R. Walsh and M. Wong, respectively.}
   \label{fig:cr_thickness}
\end{figure}

\vspace{-0.25cm}
We have utilized the \texttt{HSTCosmicrays} package to analyze $\sim76,000$ images taken with 5 different CCD imagers spanning from 1994 to present. From those $\sim76,000$, we have extracted morphological information on more than 1.2 billion cosmic rays. In Figure \ref{fig:cr_thickness}, we highlight one of the key results thus far. In each column, we show the detector thickness maps on the top row and the cosmic ray incidence maps on the bottom row. The thickness maps are generated by modeling fringe patterns present in long wavelength flat field observations \citep{malumuth_2003, walsh_2003, wong_2010}. The cosmic ray incidence maps are created by tabulating the number of times a given pixel is impacted by a cosmic ray over its lifetime.
\vspace{-0.35cm}
\vspace{-0.75cm}

%
%

\bibliography{P10-48}  


\end{document}